%% file: main.tex
\documentclass[conference]{IEEEtran}
\IEEEoverridecommandlockouts
\usepackage{cite}
\usepackage{amsmath,amssymb,amsfonts}
\usepackage{algorithmic}
\usepackage{graphicx}
\usepackage{textcomp}
\usepackage{xcolor}
\usepackage{lipsum}
\usepackage{multirow}
\usepackage{diagbox}
\usepackage{float}

\usepackage[ruled]{algorithm2e}

\def\BibTeX{{\rm B\kern-.05em{\sc i\kern-.025em b}\kern-.08em
    T\kern-.1667em\lower.7ex\hbox{E}\kern-.125emX}}
\begin{document}

    \title{Designing Short-Stage CDC-XPUFs: Balancing Reliability, Cost, and Security in IoT Devices\\

}

\author{\IEEEauthorblockN{1\textsuperscript{st} Gaoxiang Li}
\IEEEauthorblockA{\textit{Department of Computer Science} \\
\textit{Texas Tech University}\\
Lubbock, TX 79409, USA\\
email address or ORCID}
\and
\IEEEauthorblockN{2\textsuperscript{nd} Yu Zhuang}
\IEEEauthorblockA{\textit{Department of Computer Science} \\
\textit{Texas Tech University}\\
City, Country \\
email address or ORCID}
}


\maketitle

\begin{abstract}

The rapid expansion of Internet of Things (IoT) devices demands robust and resource-efficient security solutions. Physically Unclonable Functions (PUFs), which generate unique cryptographic keys from inherent hardware variations, offer a promising approach. However, traditional PUFs like Arbiter PUFs (APUFs) and XOR Arbiter PUFs (XOR-PUFs) are susceptible to machine learning (ML) and reliability-based attacks. In this study, we investigate Component-Differentially Challenged XOR-PUFs (CDC-XPUFs), a less explored variant, to address these vulnerabilities. We propose an optimized CDC-XPUF design that incorporates a pre-selection strategy to enhance reliability and introduces a novel lightweight architecture to reduce hardware overhead. Rigorous testing demonstrates that our design significantly lowers resource consumption, maintains strong resistance to ML attacks, and improves reliability, effectively mitigating reliability-based attacks. These results highlight the potential of CDC-XPUFs as a secure and efficient candidate for widespread deployment in resource-constrained IoT systems.

\end{abstract}


\input{sec1}

\input{sec2}

\input{sec3_ml_attack}

\input{sec4_reliability}

\input{sec5_selection_exp}

\input{sec6}

\input{sec7}

\section{Conclusion}

Before this work, the prevailing trend in PUF research often focused on developing entirely new PUF architectures to overcome the limitations observed in traditional designs such as APUFs and XOR-PUFs. Much literature believed that these PUF designs were inherently flawed, particularly in their vulnerability to machine learning attacks and reliability-based attacks. This led to a significant shift towards exploring alternative PUF configurations that might offer better security characteristics.

However, our study revisits the potential of XOR-PUF designs, demonstrating that they can indeed perform effectively and have considerable untapped potential when enhanced with appropriate strategies. By integrating the Component-Differentially Challenged approach and a pre-selection strategy into the XOR-PUF framework, we have shown that it is possible to significantly improve both the reliability and the ML attack resistance of these devices. This not only preserves the extensive CRP space that XOR-PUFs are valued for but also reduces their susceptibility to attacks that have undermined their utility in the past.

Our findings suggest that rather than abandoning the foundational principles of existing PUF designs, there is substantial merit in enhancing these systems with innovative modifications. The CDC approach, combined with a reliability-enhancing pre-selection strategy, revitalizes the XOR-PUF design, making it a viable option for secure and efficient implementation in resource-constrained environments like IoT devices. This work underscores the importance of adaptive innovation in cybersecurity technologies, suggesting that the evolution of existing solutions could be as valuable as the invention of new ones. By refining and augmenting established designs, we can achieve significant advancements in security technology, aligning with the evolving needs of modern digital infrastructures.

In conclusion, our research provides a way for revitalizing traditional XOR-PUF architectures, demonstrating their practical utility and potential for innovation. The enhanced CDC-XPUF model not only meets the rigorous demands of IoT security but also offers a blueprint for future research to continue exploring and improving upon existing PUF architectures. This approach fosters a more sustainable path for PUF development, emphasizing incremental innovation and the adaptation of proven technologies to meet new challenges.

\section*{Acknowledgment}

\bibliographystyle{IEEEtran}
\bibliography{cite.bib}

\vspace{12pt}

\end{document}

%% file: sec1.tex
\section{Introduction}

     The rapid expansion of Internet of Things (IoT) devices has made them integral to various industries and daily life. As these devices become ubiquitous, ensuring the security of communications within these diverse networks is important. Traditional cryptographic protocols, while robust, are often too resource-intensive for the constrained computational and power capabilities typical of IoT devices. This limitation creates a pressing need for lightweight, yet secure, authentication mechanisms tailored to the IoT landscape.

     Physically Unclonable Functions (PUFs) \cite{gassend2002controlled,gassend2002silicon,lee2004technique,suh2007physical} have emerged as a promising lightweight alternative for cryptography in such settings. By leveraging inherent manufacturing variations in integrated circuits, PUFs generate unique and physically unclonable responses to input challenges, making them ideal for device identification and authentication in IoT systems \cite{herder2014physical,becker2015pitfalls,miorandi2012internet,yu2016lockdown}. Furthermore, PUFs are generally classified into two categories: weak PUFs and strong PUFs \cite{herder2014physical}. Weak PUFs have a limited Challenge-Response Pair (CRP) space, making them suitable for cryptographic key generation. In contrast, strong PUFs like Arbiter PUFs (APUFs) and XOR Arbiter PUFs (XOR-PUFs) feature extensive CRP spaces ideal for challenge-response authentication protocols.
     
     Despite their potential, APUFs are susceptible to machine learning (ML) attacks due to their linear response behavior. XOR-PUFs were introduced to enhance ML resistance by adding non-linearity through XOR gates. However, achieving robust security with XOR-PUFs often requires increasing the XOR size, which leads to greater complexity and higher power consumption—undesirable attribute for resource-constrained IoT devices \cite{ruhrmair2013puf,ruhrmair2010modeling,ganji2015attackers,alkatheiri2017towards,alkatheiri2017experimental,aseeri2018machine,mursi2020fast,wisiolattackers}.

    Reliability is another critical concern for both APUFs and XOR-PUFs. Because of their sensitivity to environmental changes and operational variability, these devices can produce inconsistent responses, undermining practical usage and opening avenues for reliability-based attacks \cite{becker2015gap,6581579}. In such attacks, adversaries exploit the variability in responses to the same challenge under different conditions to model and predict PUF behavior. Even though some advanced PUF architectures can resist conventional ML attacks, they often remain susceptible to reliability-based attacks. While certain protocols have been developed to prevent these attacks by limiting repeated CRP queries \cite{yu2016lockdown}, implementing such measures at a system-wide level is often costly and complex. This situation underscores the need for a more fundamentally secure approach that addresses reliability at the hardware level rather than relying on external protections.

    
    These challenges underscore a gap in the current PUF design: the need for architectures that simultaneously address security vulnerabilities and the rigorous resource constraints of IoT devices. The ideal solution would enhance resistance to both ML and reliability-based attacks without suffering high hardware or energy costs.
    
    In this study, we revisit traditional XOR-PUF designs and propose an enhanced framework that overcomes these limitations by integrating two key innovations:
    
    \begin{itemize}
    \item \textbf{Implementation of a Pre-Selection Strategy:} We introduce a novel pre-selection mechanism that improves the reliability of PUF responses by filtering and utilizing only the most stable and consistent CRPs. By doing so, we significantly enhance the device's robustness in diverse and challenging environments and mitigate the effectiveness of reliability-based attacks at the hardware level.

    \item \textbf{Development of a Lightweight CDC-XPUF Design:} We propose a lightweight design based on Component-Differentially Challenged XOR-PUFs (CDC-XPUFs) \cite{yu2016lockdown,wisiol2019breaking,aseeri2018subspace,mursi2021experimental,li2022new}, a less explored variant that assigns unique challenges to each component APUF. Our design strategically reduces the number of stages within each APUF component while increasing the number of components, effectively reducing hardware overhead and power consumption. Importantly, this configuration maintains high resistance to ML attacks due to increased non-linearity and preserves the extensive CRP space necessary for effective authentication protocols.
    \end{itemize}
    
    Our research demonstrates that, with these enhancements, traditional XOR-PUF architectures have considerable untapped potential and need not be discarded in favor of entirely new designs. By refining and augmenting existing frameworks, we achieve a balance between security and efficiency that aligns with the demands of resource-constrained IoT devices.
    
    The contributions of this paper are as follows:
    \begin{itemize}

    \item We address the vulnerability to reliability-based attacks by introducing a pre-selection strategy that filters out unstable CRPs, resulting in near-perfect reliability without significant hardware modifications.
    
    \item We design a lightweight CDC-XPUF architecture that reduces hardware complexity and challenge transmission overhead, making it practical for IoT applications while maintaining a high level of security against ML attacks.
    
    \item We provide comprehensive experimental validations demonstrating the effectiveness of our approach in enhancing reliability, security, and efficiency. Our results show that the enhanced CDC-XPUF achieves a balance of performance metrics suitable for real-world deployment.
    
    \end{itemize}
    
    The remainder of this paper is organized as follows: Section 2 provides background information on PUFs and discusses related work. Sections 3 and 4 introduce the pre-selection strategy and its impact on reliability and security. Section 5 presents the design of the lightweight CDC-XPUF and details the integration of our proposed strategies. Section 6 describes the experimental setup and evaluates the performance of our design. Finally, Section 7 concludes the paper and discusses future research directions.

%% file: sec2.tex
\section{Background Information on PUFs}

To provide a foundation for the technical discussions that follow, this section describes the mechanisms of Arbiter PUFs, XOR-PUFs, CDC-XPUFs, and the types of attacks they face, including conventional machine learning (ML) modeling attacks and reliability-based attacks.

\subsection{The Arbiter PUFs}
\begin{figure}[htbp]
    \centering
    \includegraphics[width=8cm]{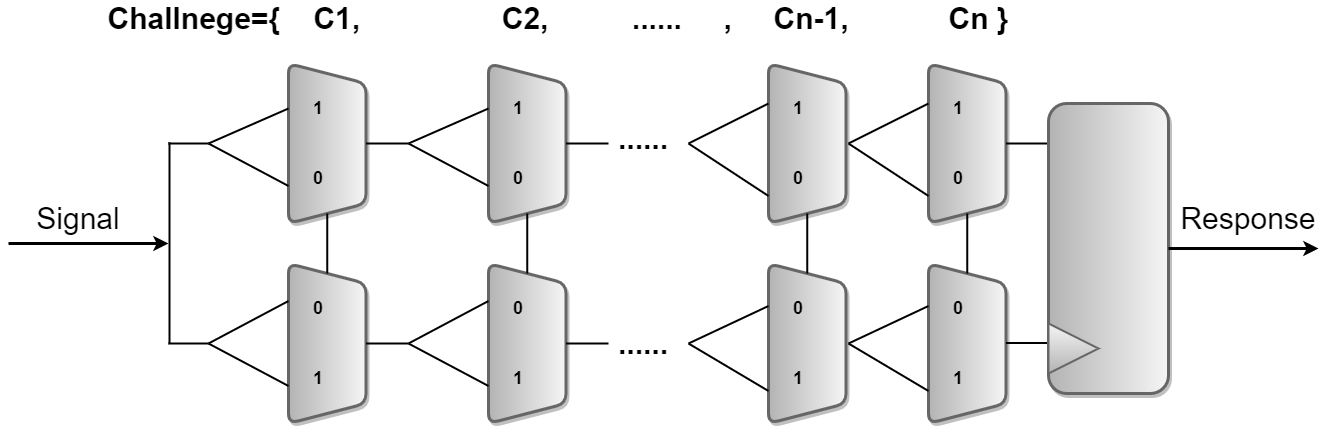}
    \caption{An arbiter PUF with n bits of challenge}
    \label{arbiter PUF}
\end{figure}

Figure \ref{arbiter PUF} showcases a simple representation of an Arbiter PUF. An $n$-bit Arbiter PUF consists of $n$ stages, each containing two multiplexers (MUXs). Upon receiving a rising signal, the signal enters the Arbiter PUF at stage one and splits into two paths. These signals traverse through gates at each stage, with their propagation paths being determined by the challenge bit supplied to the multiplexers. Ultimately, the two signals reach the D flip-flop, which acts as an arbiter. The D flip-flop ascertains whether the signal on the upper path or the lower path arrived first, returning 1 if the upper path signal arrives first, otherwise, it returns 0.

The Arbiter PUFs satisfy the additive delay model \cite{lim2004extracting}, which stipulates that the time it takes for each of the two signals to arrive at the arbiter is the summation of the delays incurred at all stages of the PUF. The response $r$ of an arbiter is defined by:
\begin{equation}
\label{equ:one}
r = \text{Sgn}(v(n) + \sum_{i=1}^{n} w(i)\phi(i)) ,
\end{equation}
where $\phi$'s are transformed challenge \cite{lim2004extracting} given by:
\begin{equation}
\label{equ:two}
\phi(i) = (2c_i-1)(2c_{i+1}-1)\cdots(2c_n-1),
\end{equation}
with $c_i$ being the challenge bit at stage $i$, $v$ and $w$'s being parameters quantifying gate delays at different stages, and $\text{Sgn}(\cdot)$ the sign function. This linear classification problem, represented by (\ref{equ:one}), makes Arbiter PUFs vulnerable to machine learning attacks \cite{ruhrmair2010modeling}.

\subsection{The XOR-PUFs}
\begin{figure}[htbp]
    \centering
    \includegraphics[width=8cm]{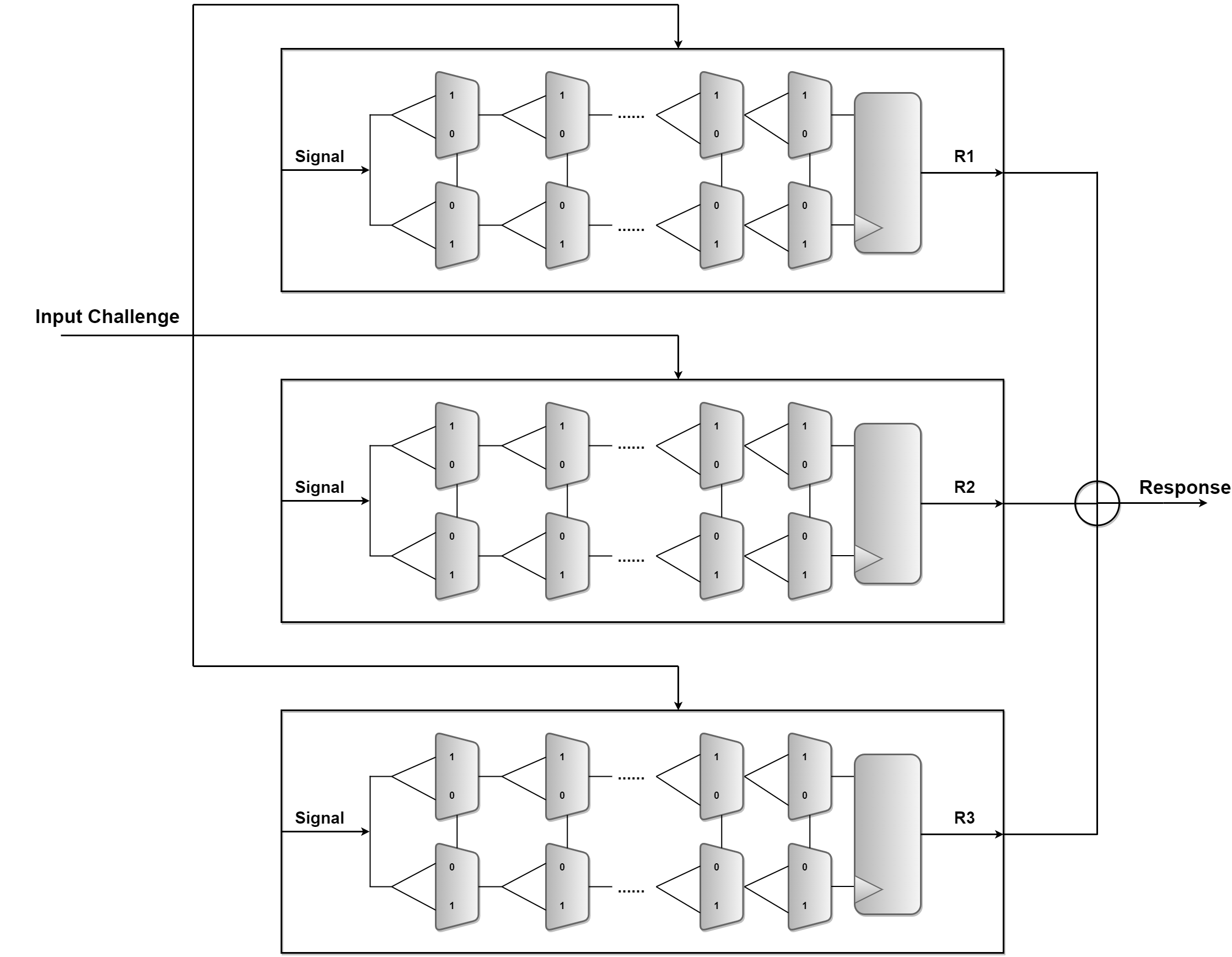}
    \caption{An XOR-PUF with 3 sub-stream and n bits of each stream}
    \label{XOR-PUF}
\end{figure}

Arbiter PUFs exhibit weak resistance against ML modeling attacks, prompting the development of XOR Arbiter PUFs, which incorporate a non-linear XOR gate with multiple Arbiter PUFs to yield the final response. This design is illustrated in Figure \ref{XOR-PUF} as an $n$-bit 3-XOR-PUF. An $n$-XOR-PUF comprises n component Arbiter PUFs, referred to as streams or sub-challenges, with the responses of all component Arbiter PUFs combined via an XOR gate to produce a single-bit response. Notably, all component Arbiter PUFs in an XOR-PUF are fed identical challenge bits.

The response of an $n$-stage $k$-XOR arbiter PUF can be expressed as:
\begin{equation}
\textbf{\emph{r}} = \bigoplus_{j = 1 \ldots k} r_{j},
\label{eq_3}
\end{equation} 
where $r_{j}$ is the internal output of the $j^{th}$ component arbiter PUF. The XOR operation increases the non-linearity of the relationship between the response $r$ and the transformed challenges $\phi$'s.

Studies have shown that XOR-PUFs achieve superior modeling attack resistance compared to Arbiter PUFs, particularly when integrated with a lockdown scheme for mutual authentication to eliminate open-access interfaces \cite{yu2016lockdown}. Nonetheless, the extension of the number of components escalates both the cost and power consumption of a PUF, a critical consideration for resource-constrained IoT devices. Moreover, increasing the number of streams can detrimentally impact the reliability of PUFs, heightening their susceptibility to reliability-based side-channel attacks \cite{becker2015gap}.

\subsection{The Component-Differentially Challenged XOR PUFs (CDC-XPUFs)}

CDC-XPUFs are an advanced variant of XOR-PUFs that enhance security by providing different challenge bits to each component Arbiter PUF. Unlike traditional XOR-PUFs, where all components receive the same challenge, CDC-XPUFs assign unique challenges to each component, increasing complexity and resistance to ML attacks. While CDC-XPUFs have been mentioned in some literature \cite{yu2016lockdown, wisiol2019breaking, aseeri2018subspace, mursi2021experimental, li2022new}, comprehensive investigations into their capabilities and limitations are still lacking.


The adoption of CDC-XPUFs in security systems, especially for IoT, encounters critical challenges that prevent their widespread deployment:

\begin{itemize}
    \item \textbf{Increased Challenge Transmission:} Each component in a CDC-XPUF receives a unique set of challenge bits, significantly increasing the total number of bits that must be transmitted to and processed by the PUF. This leads to higher power and computational demands, which conflicts with the low-power requirements typical of many IoT applications.
    \item \textbf{Vulnerability to Reliability-Based Attacks:} Despite their advanced design, CDC-XPUFs are more susceptible to reliability-based ML attacks than conventional XOR-PUFs \cite{yu2016lockdown}. These attacks exploit variability in PUF responses, analyzing them to uncover exploitable patterns \cite{becker2015gap,6581579}, thus posing a severe security risk.
\end{itemize}

%% file: sec3_ml_attack.tex
\subsection{Machine Learning Modeling Attacks}
\subsubsection{Conventional ML modeling attack}

Machine learning has become a primary tool for analyzing the security of PUFs, enabling attackers to create predictive models with high accuracy \cite{ruhrmair2010modeling}. Attackers can obtain data to build these models by eavesdropping on communications between the PUF and a server, gaining access to a large set of uniformly random challenges and their corresponding responses. This type of attack, known as a conventional CRP attack, relies solely on the PUF's CRPs and has been extensively studied.

Various ML attack methods have been proposed, including Support Vector Machines (SVM), logistic regression (LR) with resilient backpropagation (RProp), and neural networks (NNs) \cite{ruhrmair2010modeling,aseeri2018machine,santikellur2019deep,mursi2020fast}. Currently, the most effective approach for PUF modeling attacks utilizes a neural network architecture with the hyperbolic tangent (\emph{tanh}) as the hidden layer activation function and the sigmoid function as the output activation function \cite{wisiol2021neural,thapaliya2021machine}.


\subsubsection{Reliability-based ML modeling attack}
In contrast to conventional ML attacks, reliability-based ML modeling attacks require attackers to access the reliability information of the target PUF by repeatedly applying the same challenges to obtain multiple responses. This approach significantly reduces the security level of PUFs and does not rely on side-channel information such as power consumption or timing measurements, which would necessitate physical access to the PUF and complex embedding devices.

Real-world PUFs inherently exhibit some degree of unreliability due to environmental variations and electronic noise, exacerbating their susceptibility to reliability-based attacks—even when the PUF is specifically engineered to withstand conventional ML assaults. This practicality makes reliability-based attacks a serious concern. Therefore, a PUF design that only prevents conventional modeling attacks is insufficient to guarantee security; resistance to reliability-based modeling attacks is also a crucial metric for evaluating PUF designs.


There have been several proposed reliability-based ML attack methods for breaking PUF: 

CMA-ES Attack: Becker \cite{becker2015gap} introduced an attack modeling the weights of APUF components using the Covariance Matrix Adaptation Evolution Strategy (CMA-ES). While effective against simpler PUF designs, it is slow and cannot break more complex architectures like the iPUF. Gradient-Based Logistic Regression Method: Tobisch et al. \cite{tobisch2020combining} proposed an attack that simultaneously models the weights of all APUF components and their reliability information using a gradient-based logistic regression method, with constraints to avoid converging to identical components. However, this approach requires a fully differentiable model of the target PUF, limiting its applicability. Multiclass Side-Channel Attack (MSA): Liu et al. \cite{liu2022multiclass} introduced an attack employing a neural network with outputs constructed through "feature crossing." The number of output bits is determined by the product of the possible values of each output feature category, including the PUF response, a reliability measurement, and a power consumption measurement. While the method can work without power consumption data, no experimental study was conducted using only response and reliability information. Multi-Label Multi-Side-Channel Attack (MLMSA): Gao et al. \cite{gao2023mlmsa} proposed a neural network attack with different groups of output bits representing various types of information—a one-bit group for the PUF response and additional groups for different side-channel information types. Auxiliary Learning Side-Channel Attack (ALScA): The ALScA method \cite{9973338}, similar to MLMSA, uses response data, a reliability measurement, and a power measurement. However, ALScA employs different sub-networks for each group of output bits.

%% file: sec4_reliability.tex
\section{A Pre-selection Strategy to Overcome Reliability-Based Attacks}

\subsection{PUF Reliability}

Despite significant advancements aimed at enhancing ML resistance, current PUF implementations such as XOR-PUFs with large XOR sizes and CDC-XPUFs remain vulnerable to reliability-based attacks. These attacks pose a critical threat as they exploit the inherent variability in the physical properties of PUFs, leading to inconsistent outputs that can undermine security.


Although certain protocols have been developed to prevent reliability-based attacks by limiting repeated CRP queries\cite{yu2016lockdown}, implementing these protocols at a system-wide level is often costly and complex. This has prompted a need for a more fundamentally secure approach that addresses reliability at the hardware level rather than relying on external protections.

To overcome these vulnerabilities, we propose a new design strategy for PUFs that enhances reliability directly through hardware modifications. This approach begins by investigating the core of unreliability issues—specifically, the variations in delay models under different operational conditions.

\subsection{Delay Difference and PUF Reliability}

In Arbiter PUFs, the core determinant of the binary response is the delay difference $(\Delta D)$ between two competing signal paths. These paths are modulated by the binary `challenge' applied to the PUF. The delay difference $\Delta D$ is defined as the difference between the delay encountered by the signal on the upper path $(d_{\text{upper}})$ and the lower path $(d_{\text{lower}})$, formally expressed as:
\begin{equation}
\Delta D = d_{\text{upper}} - d_{\text{lower}}
\end{equation}

The binary response $(r)$ of the PUF is determined by the sign of $\Delta D$. In the absence of electronic noise and other perturbations, the response is given by:
\begin{equation}
r = \text{Sign}(\Delta D)
\end{equation}
Where:
\begin{itemize}
    \item $r = 1$ if $\Delta D < 0$ (i.e., the signal on the lower path is faster).
    \item $r = 0$ otherwise (i.e., the signal on the upper path is faster).
\end{itemize}

In practical scenarios, electronic noise and environmental variations introduce a noise term $(d_{\text{noise}})$, which affects the delay measurements. The response with noise can then be modeled as:
\begin{equation}
r = \text{Sign}(\Delta D \pm d_{\text{noise}})
\end{equation}
Where $d_{\text{noise}}$ represents the variability introduced by noise, and the `$\pm$' sign indicates that noise can either increase or decrease the measured delay difference.

A critical observation is that the reliability of the PUF response heavily relies on the magnitude of \(\Delta D\). If \(\Delta D\) is substantial, it is highly unlikely that the noise term \(d_{\text{noise}}\) will be sufficient to alter the sign of \(\Delta D\), thus preserving the integrity of the response. Conversely, when \(\Delta D\) is minimal, even slight noise can result in erroneous responses, rendering such CRPs unreliable.

One might consider employing a majority voting technique to diminish the effect of noise and increase response reliability. However, recent research \cite{li2024novelreliabilityattackphysical} has demonstrated that PUFs enhanced with majority voting can still be compromised by reliability-based attacks. Majority voting cannot fundamentally resolve unreliability issues when  \(\Delta D\) is close to zero. In such scenarios, even minor noise can flip the response, and aggregating inherently unstable responses yields marginal reliability enhancement. Consequently, PUFs enhanced with majority voting can still be vulnerable to reliability-based attacks that exploit residual variability in responses near the decision boundary. This vulnerability underscores the need for a more robust solution that addresses reliability at the hardware level.

\subsection{Pre-selection strategy to improve reliability}


Based on the observation of noise impact the reliability of PUF responses is significantly influenced by the magnitude of $\Delta D$, large values of $\Delta D$ ensure that the noise term $d_{\text{noise}}$ is unlikely to alter the outcome of $\text{Sign}(\Delta D)$, leading to stable and reliable responses. This critical observation motivates the implementation of a pre-selection strategy wherein CRPs are evaluated and selected based on the robustness of their delay differences. Only those CRPs where $\left|\Delta D\right|$ exceeds a predefined threshold are utilized, significantly enhancing the PUF's reliability by mitigating noise-induced response fluctuations. 

To operationalize the pre-selection strategy, we introduce additional delay elements into the traditional PUF architecture to selectively manipulate the delay paths based on the observed $\Delta D$ values. The following figure \ref{fig:apuf_preselection} illustrates this implementation:

\begin{figure*}[htbp]
    \centering
    \includegraphics[width=0.8\textwidth]{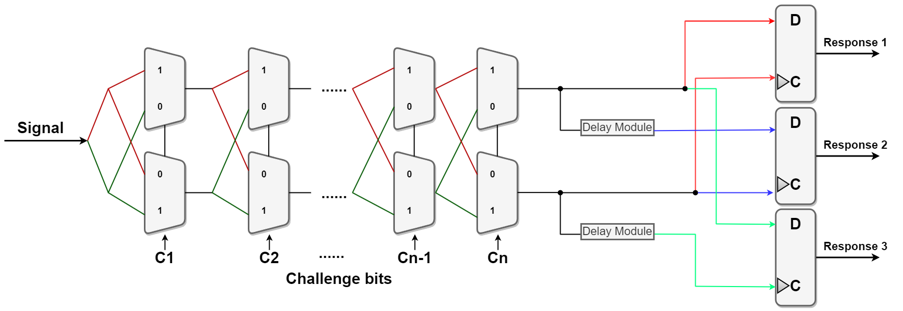}
    \caption{Implementation of the pre-selection strategy in an Arbiter PUF, illustrating the conditional path adjustments for enhancing response reliability.}
    \label{fig:apuf_preselection}
\end{figure*}

To enhance the reliability of responses, we implement a pre-selection strategy that involves evaluating three potential responses for each challenge:
\begin{itemize}
    \item \textbf{Response1:} $\text{Sign}(d_{\text{upper}} - d_{\text{lower}})$
    \item \textbf{Response2:} $\text{Sign}(d_{\text{upper}} + D_{\text{Delay module}} - d_{\text{lower}})$
    \item \textbf{Response3:} $\text{Sign}(d_{\text{upper}} - (d_{\text{lower}} + D_{\text{Delay module}})$
\end{itemize}

CRPs are selected for use only if all three responses are consistent, i.e., all are "000" or "111". This condition ensures that the absolute difference $|d_{\text{upper}} - d_{\text{lower}}|$ is greater than the "Delay module", indicating a robust delay difference that is likely to be resilient to noise impacts. This setup ensures that only the CRPs with significant delay differences, and thus higher reliability, are selected for security applications.

%% file: sec5_selection_exp.tex
\section{Pre-selection Strategy Experimental Validation and Discussion}

This section validates the effectiveness of the pre-selection strategy introduced to mitigate reliability-based attacks in PUF designs. Our experiments utilize two types of FPGA boards—Artix-7 A7-35T and NEXYS A7-50T—to demonstrate the robustness of the proposed strategy across different hardware implementations.

\subsection{Experimental Setup}

A total of 100 million CRPs were generated for each PUF instance using a pseudorandom number generator, defined by the equation:
\begin{equation}
    C_{n+1} = (a \times C_n + g) \mod m,
\end{equation}
where $C$ represents the sequence of generated numbers, $a$ is a multiplier, $g$ a constant adder, and $m$ is $2^K$, with $K$ being the number of stages. CRPs were selectively filtered to include only those producing uniform "000" or "111" responses, ensuring significant delay differences $(\Delta D)$ and enhanced reliability.

Two configurations were tested to explore their impact on PUF performance: 
1. Modules with two or four NOT gates.
2. Modules with one, two, or three AND gates.

These configurations introduced controlled delays in the signal paths, allowing the examination of their effects on PUF performance under varied conditions.

The PUF architectures were designed using VHDL within the Xilinx Vivado 15.4 HL Design Edition software suite, and implemented on FPGAs using Tool Command Language scripts for precise placement. Communication between the FPGA and the test environment was managed via the Xilinx Software Development Kit, using AXI GPIO interfaces for challenge submissions and response collections, and AXI UART for data transmission at a rate of 230,400 bits per second, facilitated by Tera Term.

Reliability tests were conducted 10,000 times per PUF instance, maintaining the operating voltage at 2.0W and ambient temperature around the Xilinx Artix\textregistered-7 FPGA at 26.0$^\circ$C, with a thermal margin of 59.0$^\circ$C (12.3W).

\subsection{Evaluation Metrics for PUF Performance}

\textbf{Uniqueness:}
Introduced by Hori et al. \cite{hori2010quantitative}, uniqueness measures the ability of PUFs to produce distinct responses across devices for the same challenges. It is quantified as:
\begin{equation}
    HU_k = \frac{4}{N_r \times N^2 }\sum_{i=1}^{N_r}\sum_{j,m=1, j\ne m}^{N} (b_{j,i} \oplus b_{m,i}),
\end{equation}
where $N$ is the number of chips, $N_r$ is the response length, and $b_{j,i}$, $b_{m,i}$ are the $i$-th response bits from the $j$-th and $m$-th PUF instances, respectively.

\textbf{Reliability (BER):}
The Bit Error Rate (BER) quantifies the consistency of PUF outputs under identical conditions and is defined as:
\begin{equation}
    \text{BER} = 1 - R = \frac{N - \sum_{i=1}^{N} (b_i == b_{ref})}{N},
\end{equation}
where a lower BER indicates higher reliability.

\textbf{Randomness:}
Randomness evaluates the balance of '0's and '1's in PUF responses to ensure unpredictability:
\begin{equation}
    p = \frac{1}{N_r} \sum_{i=1}^{N_r} b_i,
\end{equation}
\begin{equation}
    H = -\log_{2}\max(p, 1 - p),
\end{equation}
where $N_r$ is the total number of responses, and $b_i$ is a response bit.

\subsection{Experimental Results}

The experimental results are summarized in Table \ref{fig:experimental_result1}, demonstrating that the pre-selection strategy significantly enhances the reliability of the PUFs. This is evident from the substantial reduction in the BER and the maintenance of high levels of uniqueness and randomness in the responses. These improvements are indicative of the strategy's effectiveness in enhancing the operational reliability of PUFs under varied conditions.

To quantify the effectiveness of our pre-selection strategy, we define the "CRP Selection Rate" (\( R \)) as the percentage of CRPs that result in uniform responses of "000" or "111", reflecting the robustness of the response against noise and variations. This rate for APUF is calculated as follows:
\[
R = \left(\frac{\text{Number of CRPs meeting the criteria}}{\text{Total CRPs generated}}\right) \times 100\%
\]
This metric directly measures the efficacy of the pre-selection strategy in filtering out unreliable CRPs, ensuring that only the most stable and reliable responses are utilized for security applications.

Step further, configurations utilizing two delay gates emerged as particularly effective, achieving an optimal balance between securing reliability and maintaining an acceptable selection rate. Specifically, two delay gates ensured enhanced reliability of the PUF responses, with approximately 5\% of CRPs being reliably filtered. This rate is considered acceptable given that the PUFs tested are capable of providing an exponentially large CRP space, ensuring a sufficient number of CRPs remain usable for robust authentication and security applications.

Conversely, while one delay gate did increase reliability compared to no delay modification, the improvements were not substantial enough and the CRPs filtered out by one delay gate are still not reliable. The configuration with four delay gates, although potentially offering the highest level of reliability, resulted in an overly restrictive selection rate of merely 0.06\% to 0.2\% of CRPs. Such a low rate severely limits the practical usability of the PUF, as it drastically reduces the number of available CRPs to a point that may not support diverse operational requirements, especially in scenarios demanding high-frequency authentication. 

\begin{table*}[htbp]
        \centering
        \caption{Comparison of PUF performance metrics before and after applying the pre-selection strategy, featuring different delay module configurations.}
        \linespread{1.3}\selectfont
        \setlength\tabcolsep{5pt}
        \label{fig:experimental_result1}
   \begin{tabular}{|c|c|c|c|c|c|}
\hline
\textbf{PUF type} &
  \textbf{Delay Module} &
  \textbf{CRP selection rate} &
  \textbf{\begin{tabular}[c]{@{}c@{}}BER\\ (ideal value 0)\end{tabular}} &
  \textbf{\begin{tabular}[c]{@{}c@{}}Uniqueness\\ (ideal value 1)\end{tabular}} &
  \textbf{\begin{tabular}[c]{@{}c@{}}Randomness\\ (ideal value 0.5)\end{tabular}} \\ \hline
\multirow{6}{*}{\textbf{APUF 64-bit}} & \textbf{None}        & 100\%    & 0.67\%           & 42\% & 52\% \\ \cline{2-6} 
                                      & \textbf{1 AND gate}  & 70\%     & 0.02\%           & 50\% & 55\% \\ \cline{2-6} 
                                      & \textbf{2 AND gates} & 6.2\%    & \textless{}10e-8 & 52\% & 58\% \\ \cline{2-6} 
                                      & \textbf{2 NOT gates} & 5.90\%   & \textless{}10e-8 & 65\% & 56\% \\ \cline{2-6} 
                                      & \textbf{3 AND gates} & 0.06\%   & \textless{}10e-8 & -    & -    \\ \cline{2-6} 
                                      & \textbf{4 NOT gates} & 0.02\%   & \textless{}10e-8 & -    & -    \\ \hline
\multirow{2}{*}{\textbf{2 XOR-PUF}}   & \textbf{1 AND gate}  & 51\%     & 0.02\%           & 60\% & 54\% \\ \cline{2-6} 
                                      & \textbf{2 NOT gates} & 0.5\%    & \textless{}10e-8 & -    & -    \\ \hline
\multirow{2}{*}{\textbf{3 XOR-PUF}}   & \textbf{1 AND gate}  & 32\%     & 0.01\%           & 66\% & 55\% \\ \cline{2-6} 
                                      & \textbf{2 NOT gates} & 0.02\%   & \textless{}10e-8 & -    & -    \\ \hline
\multirow{2}{*}{\textbf{4 XOR-PUF}}   & \textbf{1 AND gate}  & 21\%     & 0.01\%           & 64\% & 56\% \\ \cline{2-6} 
                                      & \textbf{2 NOT gates} & 0.0004\% & \textless{}10e-8 & -    & -    \\ \hline
\multirow{2}{*}{\textbf{6 XOR-PUF}}   
& \textbf{1 AND gate}  & 2\%      & 0.008\%           & 68\% & 55\% \\ \cline{2-6} 
                                      & \textbf{2 NOT gates} & 0.0000\% & -                & -    & -    \\ \hline
\multirow{2}{*}{\textbf{10 XOR-PUF}}   
& \textbf{1 AND gate}  & 0.0000\%      & -           & - & - \\ \cline{2-6}
                                      & \textbf{2 NOT gates} & 0.0000\% & -                & -    & -    \\ \hline                           
\end{tabular}
    \end{table*}

\subsection{Discussion on Applicability to XOR-PUFs}

For XOR-PUFs, which combine the outputs from multiple Arbiter PUF components via XOR operations, the CRP Selection Rate (\( R_{\text{XOR-PUF}} \)) is determined by the intersection of reliable CRPs across all involved components. This approach is necessary because the XOR-PUF's final response is reliable only if all component PUFs simultaneously provide stable outputs ("000" or "111"). The selection rate for the XOR-PUF is thus given by the proportion of CRPs that are commonly reliable across all components:

\[
R_{\text{XOR-PUF}} = \frac{\text{Number of CRPs reliable in all components}}{\text{Total CRPs generated}}
\]

The experimental results presented in Table \ref{fig:experimental_result1} demonstrate that while the pre-selection strategy significantly enhances the reliability of APUFs and XOR-PUFs with smaller XOR sizes, the strategy becomes less effective as the XOR size increases. Due to the inherent design in XOR-PUFs, the variance in component reliability makes it difficult to apply a uniform pre-selection strategy effectively; a CRP that is reliable for one component may not be reliable for another, reducing the overall utility of the strategy. Notably, the selection rate drastically reduces for XOR-PUFs with more than two XOR gates, indicating that as XOR size increases to improve ML resistance, the applicability of the pre-selection strategy diminishes.

This presents a critical concern: PUF architectures with smaller XOR sizes, although amenable to reliability enhancements through pre-selection, are vulnerable to ML attacks. Conversely, larger XOR sizes, while resistant to ML attacks, yield a selection rate too low to be practical under the pre-selection strategy. This observation underscores the need for a PUF design that combines robust ML attack resistance with the capability to implement pre-selection reliability-enhancing strategies.

\begin{table}[h]
    \centering
    \caption{Reliability Performance of selected PUFs with different reliability enhanced strategies .}
    \label{tab:reliability_enhanced_method}
\begin{tabular}{|c|c|c|}
\hline
\textbf{PUF type} & \textbf{Reliability enhanced method} & \textbf{BER} \\ \hline
\multirow{5}{*}{\textbf{4 XOR-PUF}} & None & 2.7\% \\ \cline{2-3} 
 & selection with 1 delay model & 0.01\% \\ \cline{2-3} 
 & selection with 2 delay models & \textless{}10e-8 \\ \cline{2-3} 
 & 5 times Majority votes & 1.3\% \\ \cline{2-3} 
 & 50 times Majority votes & 0.6\% \\ \hline
\multirow{5}{*}{\textbf{6 XOR-PUF}} & None & 4.2\% \\ \cline{2-3} 
 & selection with 1 delay model & Not enough data \\ \cline{2-3} 
 & selection with 2 delay models & Not enough data \\ \cline{2-3} 
 & 5 times Majority votes & 2.0\% \\ \cline{2-3} 
 & 50 times Majority votes & 0.6\% \\ \hline
\multirow{5}{*}{\textbf{8 XOR-PUF}} & None & 5.9\% \\ \cline{2-3} 
 & selection with 1 delay model & Not enough data \\ \cline{2-3} 
 & selection with 2 delay models & Not enough data \\ \cline{2-3} 
 & 5 times Majority votes & 2.5\% \\ \cline{2-3} 
 & 50 times Majority votes & 0.7\% \\ \hline
\multirow{5}{*}{\textbf{10 XOR-PUF}} & None & 7.6\% \\ \cline{2-3} 
 & selection with 1 delay model & Not enough data \\ \cline{2-3} 
 & selection with 2 delay models & Not enough data \\ \cline{2-3} 
 & 5 times Majority votes & 3.1\% \\ \cline{2-3} 
 & 50 times Majority votes & 1.0\% \\ \hline
\multirow{5}{*}{\textbf{10 CDC-XPUF}} & None & 8.0\% \\ \cline{2-3} 
 & selection with 1 delay model &  0.005\% \\ \cline{2-3} 
 & selection with 2 delay models & \textless{}10e-8\\ \cline{2-3} 
 & 5 times Majority votes & 2.6\% \\ \cline{2-3} 
 & 50 times Majority votes & 0.9\% \\ \hline
\end{tabular}
\end{table}

\subsection{Pre-Selection Strategy for CDC-XPUF}

Given the limitations of traditional XOR-PUFs, particularly their vulnerability to ML attacks when configured with smaller XOR sizes, we explored the potential of the CDC-XPUF architecture. CDC-XPUFs present a robust architecture that supports differential challenges across multiple components, enhancing resistance to ML attacks.

The architectural complexity of CDC-XPUFs allows for more flexible management of reliability across different components. This adaptability makes them ideal for integrating our pre-selection strategy, which is designed to maintain high rates of reliable CRPs without sacrificing resistance to ML attacks.

The following table \ref{tab:cdc-xpuf-performance} presents the results from our experimental validation, showing the performance metrics of CDC-XPUFs under different delay module configurations. These results demonstrate the efficacy of the pre-selection strategy, showing substantial improvements in BER and maintaining acceptable uniqueness and randomness across various configurations of CDC-XPUFs. And with 2 NOT gates applied in the delay module, CDC-XPUFs could perform with perfect reliability.

\begin{table*}[h]
    \centering
    \caption{Performance metrics of CDC-XPUFs before and after applying the pre-selection strategy with different delay module configurations.}
    \label{tab:cdc-xpuf-performance}
    \begin{tabular}{|c|c|c|c|c|}
    \hline
    \textbf{PUF type} &
    \textbf{Delay Module} &
    \textbf{\begin{tabular}[c]{@{}c@{}}BER\\ (ideal value 0)\end{tabular}} &
    \textbf{\begin{tabular}[c]{@{}c@{}}Uniqueness\\ (ideal value 1)\end{tabular}} &
    \textbf{\begin{tabular}[c]{@{}c@{}}Randomness\\ (ideal value 0.5)\end{tabular}} \\ \hline
    \multirow{2}{*}{\textbf{CDC-2-XPUF}} & \textbf{1 AND gate}  & 0.03\%           & 58\% & 54\% \\ \cline{2-5} 
                                         & \textbf{2 NOT gates} & \textless{}10e-8 & 60\% & 55\% \\ \hline
    \multirow{2}{*}{\textbf{CDC-3-XPUF}} & \textbf{1 AND gate}  & 0.01\%           & 62\% & 52\% \\ \cline{2-5} 
                                         & \textbf{2 NOT gates} & \textless{}10e-8 & 66\% & 54\% \\ \hline
    \multirow{2}{*}{\textbf{CDC-4-XPUF}} & \textbf{1 AND gate}  & 0.009\%          & 55\% & 53\% \\ \cline{2-5} 
                                         & \textbf{2 NOT gates} & \textless{}10e-8 & 55\% & 54\% \\ \hline
    \end{tabular}
\end{table*}

\subsection{Mitigating Reliability-Based Attacks with CDC-XPUF}

The pre-selection strategy implemented in the CDC-XPUF design ensures the utilization of only robust and unique CRPs across different components. This strategy significantly reduces the risk of attackers exploiting the inherent variability and unreliability typically associated with PUF responses.

Reliability-based modeling attacks exploit the variability in PUF responses by repeatedly applying the same challenges to observe fluctuating outputs. Our strategy counters this vulnerability by selecting CRPs that demonstrate inherently stable and near-perfect reliability, effectively negating the possibility of reliability-based attacks.

Since only CRPs with consistent responses are used, the fundamental requirement for reliability-based attacks—response variability—is removed. Given the near-perfect reliability achieved, further testing specifically for resistance to reliability-based attacks may not be necessary.

By rigorously filtering out any CRPs susceptible to noise and environmental variations, the CDC-XPUF architecture ensures that the PUF responses are highly consistent. This approach not only enhances the overall reliability of the device but also drastically reduces the attack surface that could potentially be exploited by adversaries focusing on reliability weaknesses. Thus, the CDC-XPUF offers a significant advancement in PUF technology by addressing one of the most challenging vulnerabilities in traditional PUF designs.

%% file: sec6.tex
\section{Design of Lightweight CDC-XPUFs}

    The deployment of CDC-XPUFs in IoT devices faces significant challenges due to increased challenge transmission requirements and hardware demands. These issues limit the practical utility of CDC-XPUFs for resource-constrained IoT applications.
    
    Our pre-selection strategy effectively addresses the vulnerability to reliability-based attacks by filtering out unstable CRPs, significantly enhancing the reliability of PUF responses. This approach not only reduces susceptibility to such attacks but also eliminates potential reliability side-channel patterns.
    
    Building on this foundation, we focus on reducing the increased challenge transmission and hardware overhead associated with CDC-XPUFs, while maintaining strong resistance to machine learning modeling attacks. To achieve this, we examine the factors that impact the ML attack resistance of APUF-based designs.

    \subsection{Factors Impacting ML Modeling Attack Resistance of APUF-Based PUFs}

    Two primary factors influence the resistance of APUF-based PUFs to ML modeling attacks:
    
    \begin{itemize}
        \item \textbf{Number of Stages within Each Arbiter PUF Component:} Inside the arbiter PUF component, the response $r$ is determined by the additive delay model \cite{lim2004extracting}, where the total delay is the sum of delays across all stages. This model (\ref{equ:one}) represents a \textbf{linear classification} problem, making it inherently susceptible to linear ML attacks.  Increasing the number of stages can marginally enhance resistance by expanding the feature space, but the linear nature of the model remains a vulnerability
        
        \item \textbf{Number of Arbiter PUF Components (XOR Size):} Combining multiple Arbiter PUF components through XOR operations increases the non-linearity between the response $r$ and the transformed challenges $\phi$s. This added complexity significantly complicates ML models attempting to predict the PUF responses. Empirical studies have shown that increasing the number of components has a more pronounced effect on enhancing ML attack resistance compared to increasing the number of stages \cite{ruhrmair2010modeling, lim2005extracting}.

        
    \end{itemize}
    
    
    \subsection{Proposed Lightweight CDC-XPUF Design}
    
    Guided by these insights, we propose a lightweight CDC-XPUF design that balances high attack resistance with reduced hardware overhead. Our approach involves increasing the number of components while decreasing the number of stages within each component. This strategy achieves enhanced ML attack resistance with lower resource consumption, making it ideal for resource-constrained IoT environments.
    
    Key features of the proposed design include:

    \begin{itemize}
        \item \textbf{Reduced Stage Count per Component:} By decreasing the number of stages in each Arbiter PUF component, we reduce the hardware required for each component, leading to overall lower resource usage and power consumption.
        
        \item \textbf{Increased Number of Components:} Adding more Arbiter PUF components increases the non-linearity and complexity of the overall PUF response, significantly enhancing resistance to ML attacks.
        
        \item \textbf{Distinct Challenges per Component:}  Unlike traditional XOR-PUFs, CDC-XPUFs receive unique challenge inputs for each component. This dramatically enlarges the potential CRP space without necessitating longer challenge bits for each component.
        
    \end{itemize}
    
    For example, a 16-stage CDC-XPUF with 4 components can support a $2^{16^4}$ CRP space, vastly greater than the $2^{16}$ CRP space of a similarly configured XOR-PUF. This distinction allows CDC-XPUFs to benefit from reduced stage counts without sacrificing the expansiveness needed for effective identification and authentication protocols.

    By carefully balancing the number of stages and components, our lightweight CDC-XPUF design achieves enhanced ML attack resistance, reduced hardware overhead, and efficient challenge transmission.

%% file: sec7.tex
\section{Integration and Evaluation of the New lightweight CDC-XPUF Design}

\subsection{Strategic Integration of Design Innovations}

Building upon the individual strengths of the pre-selection strategy and the architectural innovations of shorter-stage CDC-XPUFs, we propose a unified design that incorporates both elements to address the dual challenges of high hardware overhead and vulnerability to ML modeling attacks. 

In theory, by applying the pre-selection strategy within the shorter-stage CDC-XPUF framework, we enhance the PUF's resistance to noise and environmental variability while maintaining a large and secure CRP space. This method ensures that despite the reduced number of stages, the security and reliability of the PUF responses are not compromised. Furthermore, the increase in the number of XOR components within each PUF component enriches the non-linearity, thereby preserving resistance against advanced ML attacks.

\subsection{Experimental Evaluation}

    \subsubsection{Evaluation Setup}
    
    To validate the efficacy of the integrated CDC-XPUF design, we conducted a series of experiments using the same FPGA board configurations as we tested in the pre-selection strategy. 
    
    In the CRP generation process, we first apply arbiter PUF components and delay module (two NOT gates) of the CDC-XPUFs on the FPGA board and utilize the pre-selection strategy to build a CRP sub-dataset. After collecting the CRP of each component, we combine these sub-datasets to build one integral dataset for one lightweight CDC-XPUF. These integral datasets are used to evaluate the PUF's performance across various metrics, including reliability, randomness,  uniqueness, hardware cost, and resistance to ML modeling attacks. 

    For ML modeling attack resistance evaluation, resilience is evaluated by employing ML attacking methods. This is executed by providing a novel training set of simulated CRPs and subsequently predicting responses to fresh challenges. In every attack, the quantity of CRPs exploited begins at a minimal level and gradually increases until a threshold (specified in the column "Training Size" of the corresponding table) is attained, which either yields a 90\% attack success rate across all twenty PUF instances or culminates in failure upon reaching 100 million CRPs. It is imperative to note that only those attacks with a testing accuracy exceeding 90\% are considered successful.

    \subsubsection{Evaluation Tools for Modeling Attack Resistance: }
    
    When evaluating the modeling attack resistance of PUFs, it is crucial that the evaluation tools possess robust attacking capabilities to make meaningful claims about the security of the PUFs under consideration. In the context of PUF modeling attacks, two methodologies have gained prominence for evaluating resistance, namely the Neural Network (NN) method \cite{mursi2020fast,wisiol2021neural} and the Logistic Regression (LR)-based method \cite{ruhrmair2013puf}. Our recent research \cite{li2022new} indicates that the adapted LR-based method outperforms the NN method in attacking traditional CDC-XPUFs with extensive stage lengths. However, in this study, we retain both methods to ensure a comprehensive evaluation of the lightweight CDC-XPUFs.
    
    To facilitate comprehension and reproductive purpose, we list the parameters for the NN attack method in Table \ref{tab0} and for the LR-based attack method in Table \ref{tab1}. 

    \begin{table}[htbp]
    \centering
    \caption{Parameters of the NN attack method for $k$-XOR-PUFs and CDC-$k$-XPUFs  }
    \linespread{1.3}\selectfont
        \setlength\tabcolsep{5pt}
    \label{tab0}
    \begin{tabular}{|c|c|c|c|c|}
    \hline
    {\textbf{Parameter}}&\textbf{Description}\\ \hline
    \textbf{Optimizing Method}                             & ADAM \\ \hline
    \textbf{Output Activation Function}                             &  Sigmoid    \\ \hline
    \textbf{Learning Rate}                 &  Adaptive      \\ \hline
                 & Layer 1 = Components $\times$ Stages  \\
       \textbf{No. Neurons in}             & Layer 2 = Components $\times$ Stages / 2\\    
       \textbf{Each Layer}& Layer 3 = Components $\times$ Stages / 2 \\
       & Layer 4 = Components $\times$ Stages \\\hline
    \textbf{Loss Function}                  &  Binary cross entropy       \\ \hline
    \textbf{Batch Size}                  &  $10^{k-1}$       \\ \hline
    \textbf{Kernel Initializer}      &Random Normal \\    \hline
    \textbf{Early Stopping}      & True, when validation accuracy is 98\% \\    \hline
    \end{tabular}
    \end{table}

    \begin{table}[htbp]
        \centering
        \caption{Parameters of the adapted LR-based attack method for $k$-XOR-PUFs and CDC-$k$-XPUFs  }
        \linespread{1.3}\selectfont
        \setlength\tabcolsep{5pt}
        \label{tab1}
        \begin{tabular}{|c|c|c|c|c|}
        \hline
        {\textbf{Parameter}}&\textbf{Description}\\ \hline
        Optimizing Method                             & ADAM \\ \hline
        Output Activation Function.                             &  Sigmoid    \\ \hline
        Base Learning Rate                 &  0.01      \\ \hline
        Loss Function                  &  Binary cross entropy       \\ \hline
        Batch Size                  &  $10^{k-1}$       \\ \hline
        Patience                  &  5       \\ \hline
        
        \end{tabular}
    \end{table}

    Although we conduct two different evaluation methods, we only list the minimum required CRP to break a certain PUF among the two methods in the later experimental result section.
    
    \subsubsection{Evaluating Hardware Cost of Lightweight CDC-XPUFs}

    When evaluating the hardware cost of PUF designs, it is essential to have a metric that provides a comprehensive and standardized measure of circuit complexity. Gate Equivalents (GE) is a widely accepted metric in the industry for this purpose. It quantifies the complexity of a digital circuit in terms of the equivalent number of basic gates, such as 2-input NAND gates, that would be needed to implement the same functionality. This makes GE a technology-independent metric that allows for meaningful comparisons across different fabrication processes.
    
    One of the primary advantages of using Gate Equivalents as a metric is that it standardizes the measurement, making it universally applicable for comparing different designs and technologies. Moreover, it offers a more holistic view of complexity than counting specific components like multiplexers and arbiters, as GE accounts for all the logic gates required in the implementation. Furthermore, as the physical size and performance of components like MUXs and Arbiters can vary with fabrication technology, using GE allows comparisons that are not biased by these variations. Additionally, having insight into the number of gate equivalents can offer a better understanding of the trade-offs between design complexity, power consumption, and performance.
    
    In this study, we adopt Gate Equivalents as the primary metric for evaluating the hardware cost of various PUF designs, including CDC-XPUF. This choice aims to present a more objective and comprehensive analysis of the hardware requirements and trade-offs involved in implementing these PUF designs.


    

\subsection{Results and Analysis}

\begin{table*}[htbp]
    \centering
    \caption{Comparison of performance metrics for lightweight CDC-XPUFs with and without pre-selection.}
    \label{fig:experimental_result2}
    \begin{tabular}{|c|c|c|c|c|c|c|}
    \hline
    \textbf{Components} & \textbf{Stages}& \textbf{BER before selection}& \textbf{Selection Rate} & \textbf{BER after selection} & \textbf{Uniqueness} & \textbf{Randomness}  \\ \hline
    8 & 16& 1.8\%& 2.4\% & \textless{}10e-8 & 68\% & 54\%  \\ \hline
    8 & 24& 4.0\%& 2.1\% & \textless{}10e-8 & 72\% & 55\%  \\ \hline
    9 & 16& 2.7\%& 1.7\%& \textless{}10e-8 & 69\% & 57\%  \\ \hline
    10 & 8& 0.3\%& 1.2\% & \textless{}10e-8 & 75\% & 59\%  \\ \hline
    \end{tabular}
\end{table*}

\begin{table*}[htbp]
    \centering
    \caption{Comparison of performance metrics for lightweight CDC-XPUFs with pre-selection strategy.}
    \label{fig:experimental_result3}
    \begin{tabular}{|c|c|c|c|}
    \hline
    \textbf{Components} & \textbf{Stages} & \textbf{Required CRP to Break} & \textbf{Success Rate} \\ \hline
    
    8 & 16  & 80m & 80\% \\ \hline
    8 & 24  & 150m & 10\% \\ \hline
    9 & 16  & Failed at 200m & - \\ \hline
    10 & 8  & Failed at 200m & - \\ \hline
    \end{tabular}
\end{table*}


The experimental validation of the lightweight CDC-XPUF design offers compelling evidence of its efficacy. The integration of a pre-selection strategy with the novel architectural design of shorter-stage CDC-XPUFs provides substantial improvements in several key performance metrics.

The unified dataset, created by combining CRP sub-datasets from each component, facilitated a detailed evaluation of the PUF's performance across various metrics—reliability, randomness, uniqueness, hardware cost, and resistance to ML modeling attacks. 

The reliability performance Bit Error Rate was consistently below \(10^{-8}\), indicating extremely high reliability. Uniqueness values averaged between 65\% to 75\%, meeting the ideal criteria for effective security implementations. Given the enormous range of potential challenges, such as $2^{8 \times 10}$ in the CDC-10-XPUF 8-bit, this percentage represents a substantial amount of unique CRPs, making CDC-XPUFs viable for security applications. Randomness metrics were also within ranges 54\%-59\%, this would be because the delays of the two selector chains are not ideally close, nevertheless equalizing the two delays is almost impracticable since the architecture of the FPGA is fixed. But in the face of an exponentially large available CRP space, the randomness performances of the lightweight CDC-XPUF is adequate for future applications.

The results highlighted in Table \ref{fig:experimental_result2}  and \ref{fig:experimental_result3} (equipped with two delay models for CRP selection) demonstrate robust resistance to ML attacks, particularly in configurations with higher component counts and reduced stages. Notably, configurations with 10 components and 8 stages resisted all modeling attacks attempted with up to 200 million CRPs, showcasing no successful breach at the 90\% attack success criterion.

    
    \begin{table*}[htbp]
        \caption{Overall hardware overhead and required transmission bits for each PUF design}
        \linespread{1.3}\selectfont
    \setlength\tabcolsep{5pt}
    \label{tab10}
    \centering
  \begin{tabular}{|c|c|c|c|c|c|c|c|}
\hline
\textbf{Components} & \textbf{Stages} & \textbf{PUF Type}                   & \textbf{Number of MUXs + Arbiters} & \textbf{GE (approximation)} & \textbf{Transmission bits} & \textbf{CRP Space} & \textbf{CRPs to break} \\ \hline
\textbf{9}          & \textbf{64}     &        \multirow{2}{*}{\textbf{XOR-PUF}}                             & \textbf{$1152+9$}  & 9292          & \textbf{64}                & \textbf{$2^{64}$}  & \textbf{40m}                   \\ \cline{1-2} \cline{4-8}
\textbf{10}          & \textbf{64}     &                                     & \textbf{$1280+10$} &  10328         & \textbf{64}                & \textbf{$2^{64}$}  & \textbf{120m}                    \\ \hline

\textbf{6}          & \textbf{64}     &    \multirow{6}{*}{\textbf{CDC-XPUF}}                                  & \textbf{$768+6$}   & 6182         & \textbf{384}               & \textbf{$2^{384}$} & \textbf{120m}                   \\ \cline{1-2} \cline{4-8} 

\textbf{7}          & \textbf{24}     &                                     & \textbf{$336+7$}   & 2732         & \textbf{168}               & \textbf{$2^{168}$} & \textbf{50m}                    \\ \cline{1-2} \cline{4-8} 
 
\textbf{8}          & \textbf{16}     &                                     & \textbf{$256+8$}  & 2112          & \textbf{128}               & \textbf{$2^{128}$} & \textbf{80m}                    \\ \cline{1-2} \cline{4-8} 
\textbf{8}          & \textbf{24}     &                                     & \textbf{$384+8$}  & 3136           & \textbf{192}               & \textbf{$2^{192}$} & \textbf{150m}              \\ \cline{1-2} \cline{4-8} 
\textbf{9}          & \textbf{16}      &                                     & \textbf{$288+9$}  & 2380           & \textbf{144}                & \textbf{$2^{144}$}  & \textbf{Over 200m}                   \\ \cline{1-2} \cline{4-8} 
\cline{4-8} 
\textbf{10}          & \textbf{8}      &                                     & \textbf{$160+10$}  & 1368            & \textbf{80}                & \textbf{$2^{80}$}  & \textbf{Over 200m}                    \\ \hline

\end{tabular}
\end{table*}

    One major disadvantage of CDC-XPUFs is the additional overhead required for transmitting a larger number of bits. Table \ref{tab10} compares the number of GEs, multiplexers (MUXs), and arbiters required, along with the number of transmission bits needed. The integration of shorter stages effectively reduces the number of transmission bits without compromising CRP space or security, aligning with the demands for low-power, high-efficiency IoT devices. GE counts are significantly lower in these configurations, simplifying circuit complexity and reducing fabrication costs.

    The experimental results support the hypothesis that combining shorter stages with the pre-selection strategy addresses the challenges of increased transmission requirements and susceptibility to attacks, meeting the stringent demands of modern IoT security frameworks.